# Embedding theory in ML toward real-time tracking of structural dynamics through hyperspectral datasets

Jonathan D Hollenbach[1], Cassandra M Pate[1], Haili Jia[2], James L Hart[1], Paulette Clancy,[1,2], Mitra L Taheri[1,3]

1 Department of Materials Science and Engineering, Johns Hopkins University, Baltimore, MD
2 Department of Chemical and Biomolecular and Engineering, Johns Hopkins University, Baltimore, MD
3 Physical Sciences Division, Pacific Northwest National Laboratory, Richland, WA

**Abstract**

In-situ Electron Energy Loss Spectroscopy (EELS) is an instrumental technique that has traditionally been used to understand how the choice of materials processing has the ability to change local structure and composition. However, more recent advances to observe and react to transient changes occurring at the ultrafast timescales that are now possible with EELS and Transmission Electron Microscopy (TEM) will require new frameworks for characterization and analysis. We describe a machine learning (ML) framework for the rapid assessment and characterization of *in operando* EELS Spectrum Images (EELS-SI) without the need for many labeled training datapoints as typically required for deep learning classification methods. By embedding computationally generated structures and experimental datasets into an equivalent latent space through Variational Autoencoders (VAE), we effectively predict the structural changes at latency scales relevant to closed-loop processing within the TEM. The framework described in this study is a critical step in enabling automated, on-the-fly synthesis and characterization which will greatly advance capabilities for materials discovery and precision engineering of functional materials at the atomic scale.

## Introduction

Two-dimensional (2D) functional materials, such as MXenes and other transitional metal-based inorganic systems, are at the forefront of materials development for improving electrodes for energy storage[1], next-generation semiconductors for electronics[2], and magneto-optics for data storage[3]. However, to leverage the advantages of these carbide, nitride and carbonitride systems, it will be necessary to understand more precisely their chemical coordination and electronic structure for a given composition and material[4,5]. Control over surface terminations in MXenes offers pathways to tune physical and chemical properties and the ability to engineer materials for target applications[5,6]. However, most characterization and discussion of termination is completed at the bulk scale using methods such as vibrational spectroscopy, nuclear magnetic resonance spectroscopy[7,8], thermogravimetric-differential scanning calorimetry[9,10], and electron dispersive spectroscopy (EDS) in the scanning electron microscopy (SEM)[11,12]. In order to probe the precise local structure of these systems, techniques with high spatial resolution are needed, such as scanning probe microscopy[13], atom probe tomography[14], and transmission electron microscopy (TEM)[15,16].

TEM offers exceptional spatial resolution, providing an information-rich characterization of material structure and local configuration. More specifically, probing electronic structure through Scanning Transmission Electron Microscopy – Electron Energy Loss Spectroscopy (STEM-EELS) reveals bonding information at or below the 1 Å level. Through Spectrum Images (SI) generated via *in-situ* STEM-EELS, scientists obtain highly spatially and temporally localized structure measurements[17,18]. This ultra-high spatial resolution, combined with significant advances in sensing technologies, such as direct detection systems[19,20], enables high rates of data collection, more than 400 spectra per second and improved signal fidelity, the ability to count electrons instead of relative intensities for indirect detection systems. While direct detection systems greatly increase the signal-to-noise ratio (SNR)[21], the rate of data collection and the size of the total dataset created[22], an approach for qualitatively and quantitatively uncovering changes in the near-edge structure, has not evolved to the same extent. Current methods rely on measuring height ratio changes, energy shifts, and relative intensities to extrapolate changes in the electronic structure in the sample[23]. This includes comparison to both labeled experimental datapoints and computationally simulated near-edge structure. With the advent of machine learning (ML), we now have access to a powerful tool for pattern recognition and distilling trends from microscopy and spectroscopy datasets[24–31]. Exploiting the power of ML tools for this purpose will be the focus of this paper.

The ability to employ ML frameworks for a rapid classification of local chemical structure on 2D functional materials would not only greatly improve our understanding of the surface functionalization of these systems but would also allow for feedback control to be added to the TEM instrument. If it were possible to embed decision-making with informed classification into the TEM, we would be able to do more than simply probe atomic structure, it would also open the door to engineering precise surface chemistries intended to maximize material performance[32,33]. For these aspirations to become a reality, the necessary capabilities of such a decision framework would need to include rapid feature extraction, classification against possible chemical configurations, and the ability to extract sufficient signal from noisy spectra with few-electron events. Recent studies have demonstrated different approaches to *some* of these capabilities, though not all[25,34,35]. In contrast, this study outlines a framework that is uniquely suited for providing rapid feature embedding and classification and enabling direct feedback control and decision for *operando* techniques such as electron beam control or heating control for engineering materials within the TEM autonomously.

Feature extraction, denoising, and background subtraction via ML has been explored in different approaches for hyperspectral datasets: Approaches include matrix factorization techniques such as principal component analysis (PCA), non-negative matrix factorization (NMF)[24,27,36], and other statistical methods as well as various "deep learning" models such as autoencoders[25,31,37,38], convolutional neural networks[28], and other tensor-based methods. Statistical methods are incredibly powerful in denoising capabilities with PCA approaches, including tools which are now incorporated directly with tools such as Digital Micrograph[26,39,40] and Hyperspy[41]. Matrix factorization has also been a proven method for deconvolution of overlapping edges to map SIs[24] and extracting distinct spectral features for improved spatial mapping of compounds[27,36]. Deep learning approaches are similarly good performers in denoising and classification of spectral data. Approaches such as random forest[35,42] and CNN offer high classification accuracy, but necessitate large, labeled datasets[28,29]. Autoencoder approaches, including variation autoencoders (VAEs) have demonstrated both denoising[37] and the creation of non-linear feature embedding for clustering and classification[25]. The dimensionality reduction in VAEs differs from matrix factorization methods due to the non-linear nature[43] and the ability to decouple more significant factors without dataset bias unlike PCA and similar methods[44]. Most importantly, the latent embeddings from VAEs have been demonstrated to relate multiple features across different inputs, such as relating STEM imaging to spectral data[38], or spectral features to structural descriptors[25,34].

Of the classification methods described above, a common challenge is the reliance on extensive, labeled training data. By this we mean the use of datasets of any modality where the structure or bond interactions are known and classified for each datapoint. This can be done through traditional analysis by a human or by using computational means to generate simulated datasets wherein the structure calculation is known. The former is not feasible in many cases, and the latter can be prohibitively computationally expensive. For spectroscopic datasets such as EELS, it is common that not all present structures and potential intermediates are known, and these methods are not designed for classifying unknown labels. Additionally, the number of datapoints for these transient states may not be significant enough to form distinct clusters in unsupervised methods, as demonstrated in the previous study on $SrFeO_3$[37]. For a closed loop framework to classify *operando* experiments, a framework cannot rely on the availability of large, labeled datasets, computationally or experimentally generated, and needs to provide a robust unsupervised feature extraction capable of isolating and classifying chemical configurations.

In this paper, we demonstrate a variational autoencoder framework designed for embedding theory for the classification of highly spatiotemporal localized STEM-EELS datasets, which will provide the basis for precise feedback loops and control in future autonomous studies. The unsupervised encoder model acts to embed both low signal-to-noise ratio (SNR,) single-pixel spectra and computationally simulated "fingerprint" spectra into the same latent feature representation. Fingerprint spectra act as a representative datapoint used as the basis of classification for a given label corresponding to different structural compositions. By resolving the clustering behavior of experimental data distribution against the feature space based on computationally derived structures, the model is capable of estimating classification and uncertainty in a semi-supervised approach. The result is a rapid classification of noisy single spatio-temporal pixels against a library of a few simulated structures. This approach removes the need for large labeled experimental or theoretical datasets in addition to providing an unsupervised, non-linear latent space with informed distributions of simulated structures.

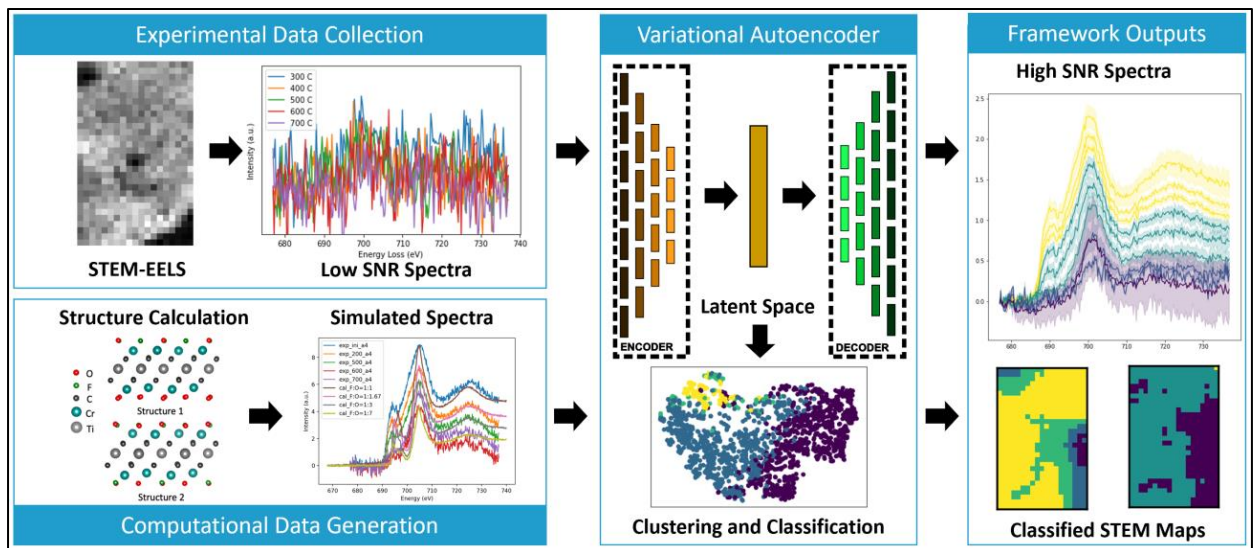

**Figure 1**: Diagram of the variational autoencoder, describing how experimental data are used to train unsupervised feature extraction, whereas computational data are used for classification and cluster analysis resulting in denoised signals and classification mapping.

## Results

To demonstrate the framework discussed above, an *in-situ* EELS-SI dataset was used consisting of three scan areas of $Cr_2TiC_2O_xF_y$ annealed up to 800° C. The full dataset includes 24 EELS-SI consisting of 570 individual spectra, each measured at a wide dispersion in order to include Ti $L_{2,3}$, O K, Cr $L_{2,3}$, and F K edges within the measured window. This experimental dataset was previously discussed in Hart *et al*.[15] but the SNR on single-pixel spectra was too low to resolve any spatial relevant changes to the near-edge structure. This necessitates the use of machine learning to probe both spatial and temporal accuracy of the 'as collected' dataset without the need to integrate along either axis.

Four edges are captured within the energy window, the Ti $L_{2,3}$, O K, Cr $L_{2,3}$, and F K edges. The change due to annealing is most notable in the O K-edge and F K-edge, shown in **Figure 2**, since the carbon, chromium, and titanium backbone is not expected to change significantly, whereas the surface terminations in the form of water, oxygen and fluorine are removed at different annealing temperatures. The surface functionalization changes are most visible in the F K-edge throughout the annealing process, with a direct relationship between the first peak height and annealing temperature. Isolating this shoulder-peak ratio returns a characteristic functionalization quantification that can be used to indicate changes across the entire dataset. If no shoulder remains, but a peak still exists, this will indicate no remaining fluorine terminations, with the remaining peak being the result of Cr $L_1$-edge and aluminum fluoride, which was noted in the previous study[15]. These structural changes, present in the experimental datasets, are supported by simulated spectra based on computed structures for various chemical configurations. These compositions were chosen based on the EXELFS (Extended Energy Loss Fine Structure) analysis in a prior study using the Cr K-edge for this material system[45]. Multiple configurations were simulated for each composition based on density functional theory (DFT) structure calculations performed at different ratios of F:O.

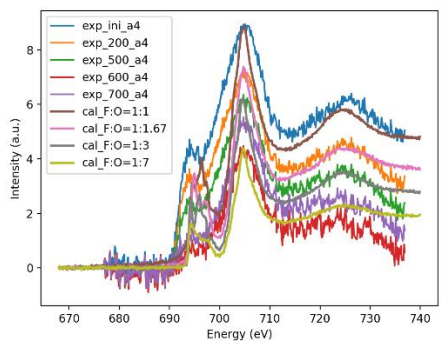

**Figure 2**: Spatially averaged spectra for given annealing temperatures ranging from initial room temperature (25°C) to 700°C, as indicated on the inset as experimental ("exp") data. Corresponding simulated spectra derived from DFT-generated structures shown as solid lines as denoted in the inset as "cal" (DFT-generated) curves for different ratios of F:O, ranging from 1.1 to 1.7, as indicated.

The temporal temperature series demonstrates a conventional analysis wherein spatial information is lost in favor of increasing the SNR. Averaging across the spatial axis increases the signal as demonstrated in **Figure 3a-d**, indicating that it is extremely important to average the signal over multiple pixels (here shown up to 32-pixel averaging). In stark contrast, ML methods, including PCA and deep learning approaches, are powerful approaches for extracting signal, *even at the individual pixel level*, increasing the spatio-temporal resolution to match the maximum rate of the detector of 400 frames per second[19]. Figure 3e,f illustrates this capability, where the SNR is comparable to the spatially averaged spectra. This demonstrates that ML means can act as an effective alternative to averaging datapoints across spatial or temporal axes in order to increase signal above the noise variation intensity.

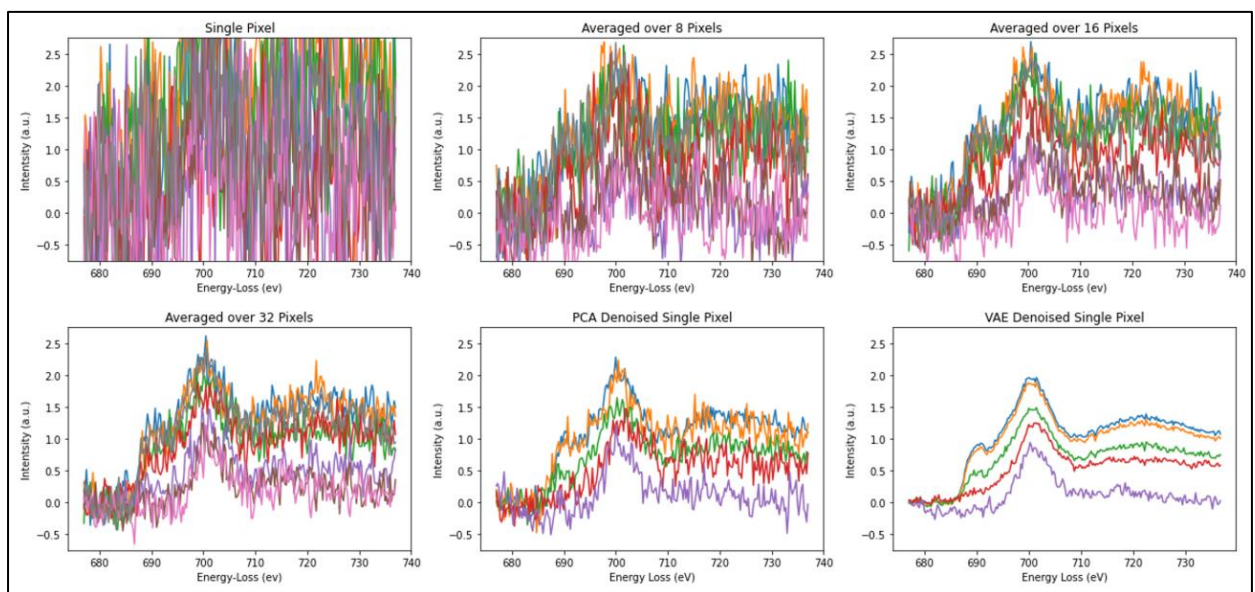

**Figure 3**: Intensity as a function of energy loss for: (a-d) single pixel (a), a few pixels (8- and 16- in (b) and (c)), versus many (32-) pixel averages for the dataset (d). Images (e, f) denote single pixel results as denoised by PCA (e) and VAE (f) approaches.

Single-pixel spectra are used to train a variational autoencoder, creating a latent space fitted to a normal distribution "prior." The simulated spectra are then fed into the trained encoder, giving a mean and log-variance as the "fingerprint" for each computationally derived structure. Sampling this mean and log-variance with a random normal density function creates a distribution for each composition which acts as the classification label. The distribution generated from the encoder is learned from the experimental dataset in the unsupervised training dataset, and decoding these latent distributions illustrates how these learned features are represented in the original near-edge spectra. The generated latent space datapoints serve as the training data for the latent space semi-supervised classifier. Applying this trained classifier to the experimental embeddings results in direct label classification for single-pixel results.

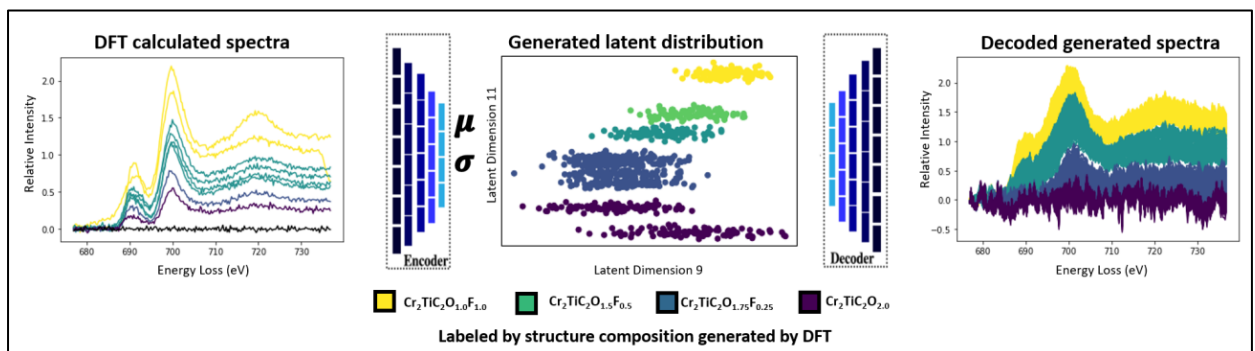

**Figure 4:** Workflow to illustrate the creation of a labeled latent distribution training dataset based on DFT-generated structure calculations (left) with distribution learning from the encoder trained on experimental data, producing decoded generated spectra (right).

A Gaussian process classifier was used for this task since the training dataset has already been generated from a normal distribution. An additional advantage of this classifier is the ability to extract probabilities for identifying datapoints with high uncertainty in prediction. Applying the trained classifier to the latent embeddings of the experimental data generates a classification which can be constructed into a full STEM map. The results can be seen in **Figure 5,** as produced by the semi-supervised classifier run on the latent encodings of single pixel data (top row). This is compared with a classification map similarly generated by denoising each spectrum via PCA (bottom row) and assigning an estimated structure classification based on the relative peak and shoulder heights, as indicated in Figure 2.

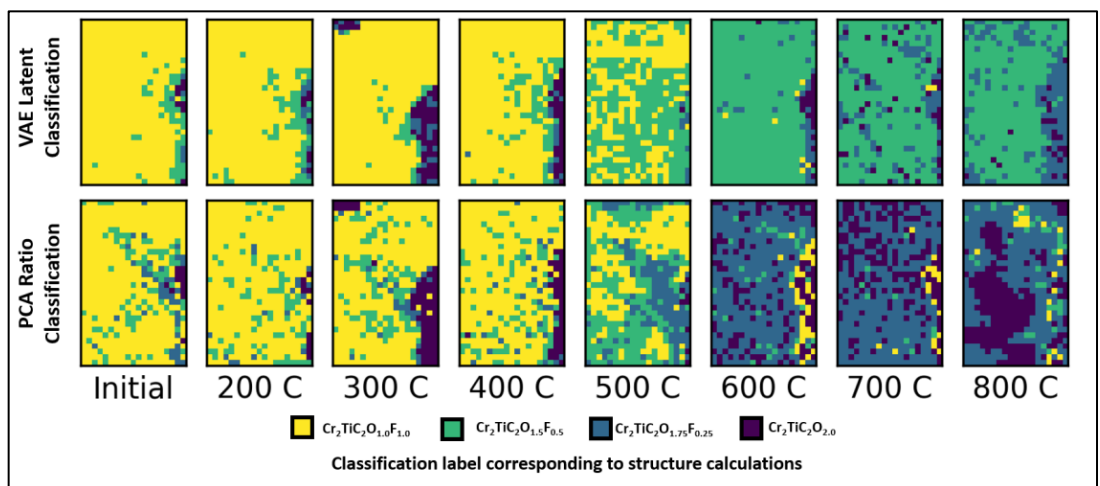

**Figure 5:** STEM classification as produced from both the VAE semi-supervised approach (top row) and peak-shoulder ratio measurements on PCA-denoised results (bottom row) across each annealing temperature studied (from RT to 800 C). Color key as indicated.

The classification results for both the VAE approach and the PCA approach readily identify reduction in fluorine content, dominant at 600°C, as denoted by the color change from predominantly yellow (50% F) to predominantly blue (indicating 12.5% F) for PCA results, and from yellow to green (25% F) using the VAE approach. This supports previous studies[15,45] on this dataset identifying the change in near-edge structure of the F K-edge at high temperatures and the reduction in fluorine nearest neighbor peaks in the Cr K-edge extended fine structure. The two approaches thus differ greatly in classifications at high annealing temperatures. A major difference occurs at the 600°C map along the right side of the STEM-EELS map where the PCA classification maps result in high fluorine content as compared to the VAE results that indicate a far less striking reduction in F content. At lower annealing temperatures, this same region of the sample is low in fluorine, as is evident in both approaches. The variance in low temperature classification maps is higher in the PCA approach.

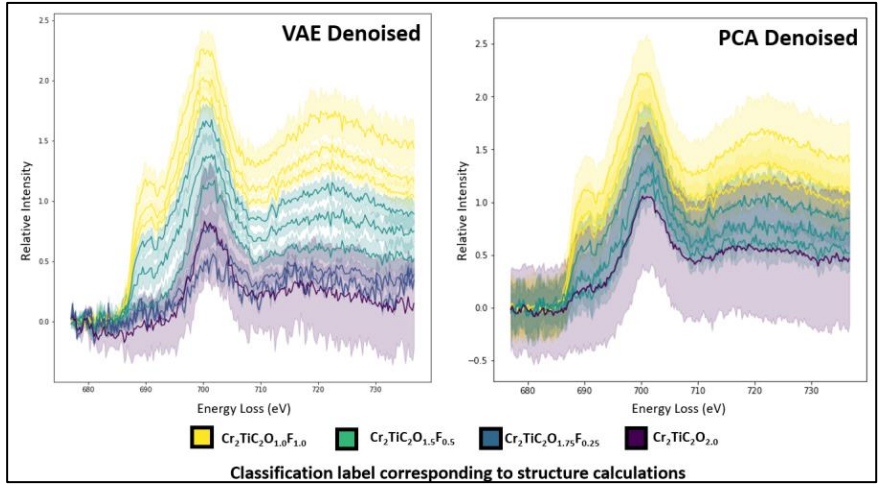

**Figure 6:** Denoised F K-edge spectra via VAE (left) and PCA (right) approaches shown as solid lines, with means (solid lines) and standard deviations (lightly shaded regions) for each classification label based on the input computationally generated spectra. Color key as given below the curves for different amounts of fluorine from 50% (yellow) to 0% (purple).

Classification maps illustrate the trends in structure changes with temperature, but further analysis of the denoised EELS spectra will be needed to fully evaluate the changes present in the dataset. The denoised signals produced by both PCA and VAE approaches can be seen in **Figure 6**. Notably, the VAE-denoised signal depicts significant changes in the annealed spectra with minimal to no fluorine content compared to the PCA-denoised approach. The residual shoulder present in the PCA data indicates a statistical bias present in the data as the annealed spectra are known (from a previous study[15]) to produce a single peak. Importantly, the pre-peak standard deviations are considerably higher in the PCA-denoised results compared to the VAE-denoised ones. This comparison shows that the VAE approach is much better equipped to extract the non-linear features of the near-edge structure signatures, which is further supported by **Figure 7a** comparing the feature space embeddings. By comparing the dimensionality reduced space, reduced to two dimensions via t-distributed stochastic neighbor embeddings (t-SNE)[46], it is evident how structural changes present in the dataset are directly related to extracted features in the latent space representations of the VAE. PCA is still capable of extracting trends in the dataset, evident by the gradient of labels across the embedding axes, but the clustering capabilities in the VAE produce more cohesive regions of each structure class. This is further supported by the silhouette score in **Figure 7b** indicating that, for each increase in the temperature series, the VAE outperforms the PCA approach.

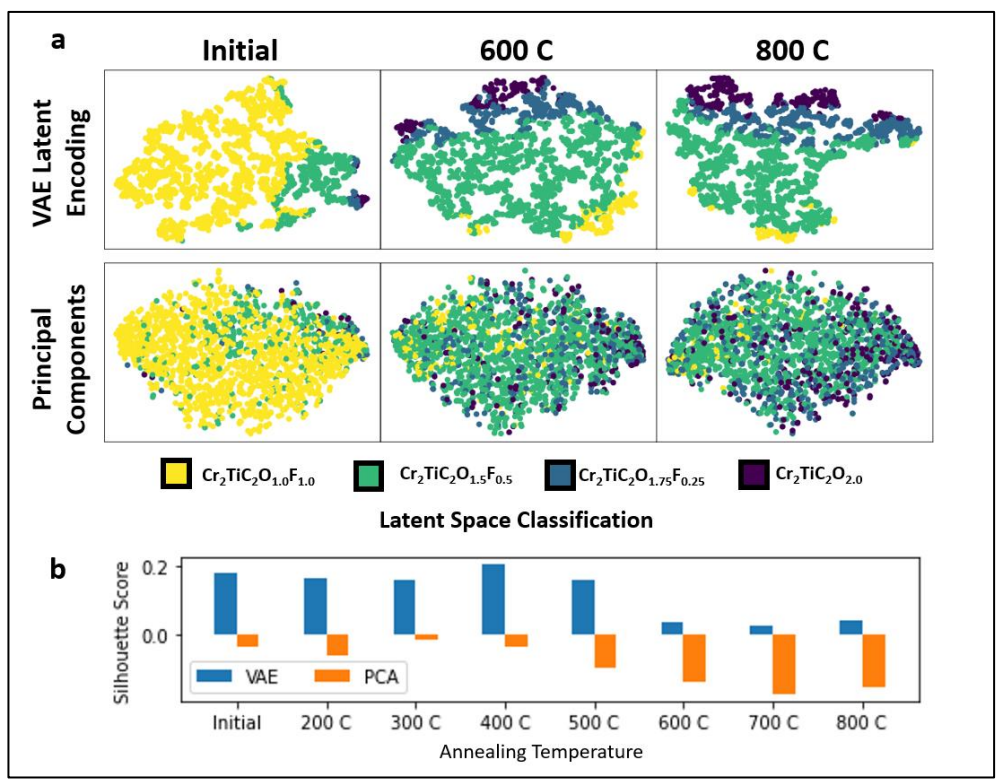

**Figure 7:** (a) t-Stochastic Neighbor Embedding (t-SNE) plots for both VAE latent encodings and principal components, demonstrating how the feature extraction of the latent encodings is more directly related to structural features present in the EELS dataset compared to PCA. This is supported quantitatively by the silhouette score (b), which is a measure of clustering performance given by the spatial distance of 'like' datapoints compared to 'unlike' datapoints.

## Discussion

This study demonstrates the capabilities of variational autoencoders to decode structural changes across an EELS dataset and establish the correlation between these latent variances and computationally generated structures. This framework does not act as a replacement for a thorough *ex-situ* comparison and careful quantitative analysis. Instead, this VAE framework enables the next step towards intelligent microscopy by providing rapid classification and a latent feature-embedding informed in the near-edge structural changes present in the EELS dataset. Leveraging these capabilities will allow for feedback loops on a temporal scale relevant to chemical changes within the microscope, thereby allowing *operando* control of chemical functionalization.

The requirements set out for enabling intelligent microscopy were as follows: rapid extraction of relevant chemical features, classification against possible chemical configurations, and inference on low SNR datasets. Unlike existing solutions, this framework does not rely on large, comprehensive, labeled ground truth data for classifier training. Instead, it relies only on the availability of a few hand-picked structure calculations for the basis of classification. This differs from other dominant methods such as convolutional neural networks (CNNs) or random forest models, which rely on training data consisting of each potential class label. Additionally, this framework differs from other unsupervised clustering algorithms as it introduces a means to establish a basis for classification within the model framework without a "human in the loop" interpretation of clusters. Finally, the ability to perform inference on noisy single pixel datapoints is on par with the denoising performance of popular, accepted methods such as PCA, but the VAE latent representation outperforms the dimensionality reduced feature space of principal components in its ability to extract features, which directly corresponds to structural changes present in the EELS dataset.

Several challenges are still present in this study and will need to be addressed going forward. Pushing the limit on spatio-temporal resolution of EELS results in difficulty of identifying valid ground truth labels for validation of classification results. Previous studies have relied on integration of multiple spatial or time points within the dataset to provide a basis for validating the machine learning results. However, if variation across each axis is high, this results in dilution of sparsely represented features, which can result in ground truths that are not representative of the original dataset. For the MXene annealing data presented in this study, the best available ground truth comes from spatially averaged or single-pixel PCA-denoised results. For validating a framework targeted at moving beyond those limitations, a reporting accuracy based upon the aforementioned ground truths would not be representative.

The framework presented here offers a necessary first step towards enabling engineered functional materials to be created with atomic precision. This framework is focused on exploring the chemical space present within STEM-EELS data and doing so with high spatial and temporal precision. This ability to extract signal from small volumes and time steps in a method that can be automated will provide the foundation for building real-time feedback loops into the TEM. Estimating candidate structures in real-time shifts the microscope from a characterization tool into a processing tool, capable of targeting local structure with the precision of the convergent electron beam. Using classification maps with precision electron patterning toolsets[47], inducing structural changes in response to local real-time classification will enable precision engineering of materials. Before such methods can be employed, however, there are several additional steps, beyond what we have achieved here, that will need to be implemented. For example, paired with any classification estimate, an uncertainty quantity should be included to provide context to further models regarding the error and extrapolation from the VAE framework. Two sources of uncertainty are propagated into each prediction, from the latent encoding and from the classification method. Any decision framework utilizing inference from this framework will need to weight inferences based on the paired uncertainty. Additionally, including this uncertainty component will aid in active learning methods such as Bayesian frameworks[48], critical for exploratory closed-loop methods.

Furthermore, the present simulation results rely on the availability of pristine crystal structures, which will not account for defects that could arise during the annealing process. Unfortunately, due to the complexity of structural changes and the number of possible defective configurations, structures with defects were not considered for the training dataset in this work. Encouraged by the success of our introduction of oxygen vacancies into the system, the use of refined (more realistic) structures can enhance the agreement between simulation and experiments, and thus further improve the accuracy of the classification. By combining DFT calculations for structures that include defects, along with the uncertainty evaluation and active learning techniques mentioned earlier, we can effectively and efficiently identify actual structures without the need to construct an extensive DFT dataset in a high-throughput manner. Ultimately, this approach paves the way for the realization of automated, on-the-fly synthesis and characterization—an exciting prospect that could elevate material design to the next level.

## Methods

Model Structure

The LogicalEELS framework consists of a variational autoencoder (VAE) trained to extract features from high SNR experimental data, and a secondary standard encoder used to relate the low SNR data to the same latent feature space. The VAE for unsupervised feature-extraction is based on the previous RapidEELS autoencoder framework[37] where 1D convolutional layers are stacked, with a single dense layer performing the last bottleneck step. Unlike traditional autoencoder frameworks, VAEs utilize a Kullback–Leibler (KL)[49] divergence factor to fit the latent variables to a probability distribution such as a Gaussian Normal[43,50]. The main variational encoder-decoder pair is trained to process high SNR experimental data such as spatially averaged, time-integrated, or PCA-denoised data to build out the feature embeddings. The secondary encoder is given a loss function related to the difference between the latent outputs of low SNR encoder and the latent mean outputs from the high SNR encoder for a given raw and denoised training datapoint. Through this structure, both noisy, single pixel datapoints and high SNR-denoised or simulated EELS spectra can be related in a single, continuous latent feature space.

The additional capability of this structure is the ability to use the variational encoder to return a distribution of latent embeddings sampled from the mean and log-variance outputs. This distribution is directly related to the learned structure from the experimental dataset used to train the VAE in an unsupervised manner. By doing so, it is possible to generate a labeled dataset based on computed structures for training a supervised clustering algorithm without the need for running many different variations of structure calculations and simulating near-edge structure spectra for each.

Model Setup and Training

As discussed in Pate *et al.*, the mean absolute error is used for the reconstruction loss as it is more resistant to outliers than the mean squared error. The reason PCA-denoised data is used in place of spectra summed over an area is due to spatial changes in the spectra across the collection area. This variance is due to heterogeneity in the sample thickness more than functionalization changes at a given condition and hence gives rise to noise in the analysis.

TensorFlow 2.11.0[51] and the Keras API was used to construct and train the encoder, decoder and neural network classifier models. Batch and layer normalization were avoided due to consistent tendency for overfitting the datasets. A test-train split of 10% to 90% was used for training the VAE as a method of validating whether new data can be embedded into the latent space without falling out of domain.

Latent Space Classification

The feature embeddings extracted from the dataset in the VAE latent space are trained through unsupervised means. Unsupervised clustering algorithms can be employed at this point to pair similar features into groupings based on the dimensionally reduced space. Instead, this study employed a supervised classification algorithm trained on the latent space features generated from the simulated spectra. The labels directly correspond to chemical compositions used as the basis for the structure calculations. We tested several classification algorithms but the Gaussian process classifier[52], as implemented in Sci-Kit Learn[53], was chosen due to the normal distribution of the training data.

Since low-latency is a primary focus of the framework, an additional classifier can be included by constructing a dense neural classifier and appending this to the experimental encoder. This method was introduced in the previous study[37] for classification and consists of a few dense layers applied at the end of the encoder layers. This is trained using the same latent space dataset generated from the simulated dataset. This effectively forms a convolutional neural network consisting of the encoder paired with the dense classifier which has the advantage of being accelerated using edge compute devices such as Field Programmable Gate Arrays[54]. This is unlike the algorithmic based classification used in the Gaussian process classifier. The result is a reduction in the overall framework latency at the expense of a small trade-off in accuracy. A full comparison breakdown of each classification accuracy test can be found in the Supplementary Information.

*In-situ* Electron Spectroscopy
The EELS Spectrum Images used for this study were collected using a JEOL 2100F, equipped with a Gatan Imaging Filter and a K2 Summit Direct Detection Sensor. Pixel size was about 13 *nm* with an exposure time of about 0.1*s* at an energy dispersion of 0.125 eV per channel. The MXene samples were deposited onto a DENSsolutions Lightning D9+ sample holder for heating and biasing Eight different conditions were measured, including the initial condition, at each of seven temperatures up to 800°C in increments of 100°C. Data used for this study were previously published in Hart *et al.*[15] and this study serves to add additional context to the spatial component of the experiment through the aid of machine learning. Additional information on sample preparation and microscope configuration can be found in the Methods section and in the Supplemental Information.

Dataset Processing
The EELS dataset was imported from Gatan DM4 files as Spectrum Images, using Hyperspy's Python API[41]. No spatial information from the dataset is passed into the autoencoder, only individual spectra. For the input dataset, spectra were rebinned, from a 0.125 eV energy dispersion to a 0.3 eV one, in order to match the measured energy resolution of the electron source. Additional pre-processing aligned all spectra in the energy axis to the Cr-L2 and Cr-L3 edges, and normalized intensities of the same peaks. A set window of 677 to 737 eV was extracted in order to capture the F K-edge. A power-law fit for background subtraction was used to isolate the F K-edge from the background. No deconvolution was performed with the zero-loss SI since only the high-loss signal was collected at each temperature and using the zero-loss peak from before heating would not account for any thickness changes occurred throughout the annealing process. Dual EELS mode, collecting the zero-loss and high-loss signals simultaneously is not feasible on the K2 Summit due to the small sensor size. The analysis was performed on both the 'as collected' spectrum images, as well as a PCA-denoised dataset generated by a Digital Micrograph[26] using 10 principal components.

Simulating EELS Spectra

Plane-wave periodic density functional theory (DFT) calculations for the MXene structures, including the lattice constants and atomic positions, were carried out using QUANTUM ESPRESSO[55]. PBEsol[56] was chosen as the exchange-correlation functional, as it was found to reproduce the experimental formation enthalpy adequately[57]. The convergence criteria were set at $1\times10^{-8}$ eV for total energy per unit cell and 0.01eV/ Å for forces. The plane-wave cutoff energy was set at 550 eV and a *k*-point grid of 10×10×1 was chosen according to the convergence tests (within 0.01 eV/atom). The two surface layers contained eight atoms in total to enable modeling of the surface-termination changes during the annealing process. Specifically, we considered full O termination (O:F = 8:0) and three mixed terminations (O:F = 7:1, 6:2 and 4:4). A vacuum region of 15 Å was used on both surface layers.

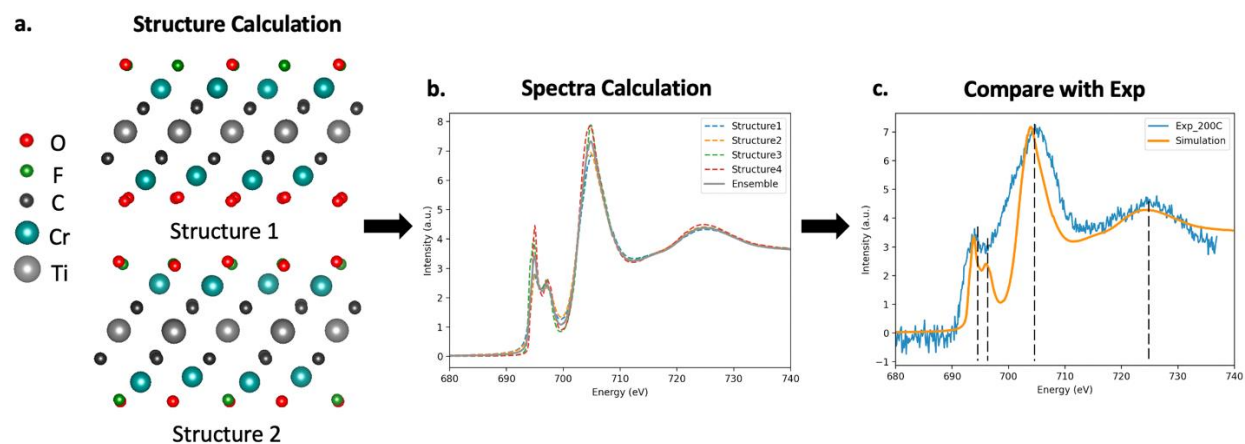

**Figure 8:** Simulation workflow. (a) Step 1: Calculate possible configurations for a user-supplied composition. Two example stable structures are shown for a relaxed MXene with a fluorine-to-oxygen ratio, F:O of 3:5. (b) Step 2: Calculate spectra for each stable structure and obtain a representative ensemble average (here four structures were used). (c) Step 3: Comparison with experimental results and fine-tuning of broadening parameters, if needed.

Core-loss spectra calculations were carried out using Finite Difference Method Near Edge Structure (FDMNES)[58]. We compared the spectra calculated based on DFT to those from time-dependent DFT (TDDFT); the differences between the resulting spectra are inconsequential. Quadrupolar components and relativism effects were also considered. Based on the convergence test evaluated by cosine similarity and Pearson correlation (0.99), the cluster radius was set as 6 Å and the cell size was set as 2×2. For each composition (different ratio between fluorine and oxygen), the spectra were calculated from the site-averaged O/F-K edge of all energy preferred configurations.

## Data Availability
Access to the full EELS dataset is available upon request from the corresponding author. Sample spectra are provided in the Github repository for this project.

## Code Availability
Python code for this study can be found at https://github.com/hollejd1/logicalEELS containing the autoencoder framework, notebooks used for training, and a few sample spectra for classification.

**Conflict of Interest Statement:** All authors declare no financial or non-financial competing interests.

**Acknowledgements:** MLT acknowledges funding in part from US Department of Energy, Office of Basic Energy Sciences through contract DESC0020314, in part from the Office of Naval Research Multidisciplinary University Research Initiative (MURI) program through contract N00014-201-2368. JH and MLT acknowledge funding in part from UES on behalf of the Air Force Research Laboratory under contract S-111-085-001. This research was supported by the AT SCALE Initiative at Pacific Northwest National Laboratory (PNNL). PNNL is a multi-program national laboratory operated for the U.S. Department of Energy (DOE) by Battelle Memorial Institute under Contract No. DE-AC05-76RL01830. Computing resources for HJ and PC were provided by the Advanced Research Computing at Hopkins (ARCH) high-performance computing facilities, which is supported by National Science Foundation (NSF) grant number OAC 1920103.